\documentclass[aps,twocolumn,groupedaddress,showpacs,showkeys,floatfix,dvipdfmx,nofootinbib]{revtex4-1}

\usepackage{wrapfig}
\usepackage{graphicx}
\usepackage{colordvi}
\usepackage{color}
\usepackage{amsmath}
\usepackage{amssymb}
 \usepackage{comment}
\usepackage{natbib}
\usepackage[normalem]{ulem}
\usepackage[small, bf]{caption}
\usepackage{tikz}

\newcommand{\ve}[1]{\mbox{\boldmath $#1$}}

\begin{document}
\title{Rare-Event Sampling Analysis Uncovers the Fitness Landscape of the Genetic Code  }
\author{Yuji Omachi *}
\affiliation{
Graduate School of Sciences
The University of Tokyo, 7-3-1 Hongo, Tokyo, 113-0033, Japan.}

\author{Nen Saito  *\dag }
\affiliation{
Graduate School of Integrated Sciences for Life, Hiroshima University, 1-3-1 Kagamiyama, Higashi-Hiroshima City, Hiroshima, 739-8528, Japan;
Exploratory Research Center on Life and Living Systems, National Institutes of Natural Sciences, 5-1 Higashiyama, Myodaiji-cho, Okazaki, Aichi
444-8787, Japan;
Universal Biology Institute, The University of Tokyo, 7-3-1 Hongo, Tokyo, 113-0033, Japan. }

\author{Chikara Furusawa\dag}
\affiliation{
Graduate School of Sciences
The University of Tokyo, 7-3-1 Hongo, Tokyo, 113-0033, Japan;
RIKEN Center for Biosystems Dynamics Research, 6-2-3 Furuedai, Suita, Osaka, 565-0874, Japan;
Universal Biology Institute, The University of Tokyo, 7-3-1 Hongo, Tokyo, 113-0033, Japan. 
}
\footnotetext[0]{\footnotesize { * These authors contributed equally to this work.\\
\dag corresponding authors; nensaito@hiroshima-u.ac.jp; \\chikara.furusawa@riken.jp}}


\bigskip

\begin{abstract}
The genetic code refers to a rule that maps 64 codons to 20 amino acids. Nearly all organisms, with few exceptions, share the same genetic code, the standard genetic code (SGC). While it remains unclear why this universal code has arisen and been maintained during evolution, it may have been preserved under selection pressure. Theoretical studies comparing the SGC and numerically created hypothetical random genetic codes have suggested that the SGC has been subject to strong selection pressure for being robust against translation errors. However, these prior studies have searched for random genetic codes in only a small subspace of the possible code space due to limitations in computation time. Thus, how the genetic code has evolved, and the characteristics of the genetic code fitness landscape, remain unclear. By applying multicanonical Monte Carlo, an efficient rare-event sampling method, we efficiently sampled random codes from a much broader random ensemble of genetic codes than in previous studies, estimating that only one out of every $10^{20}$ random codes is more robust than the SGC. This estimate is significantly smaller than the previous estimate, one in a million. We also characterized the fitness landscape of the genetic code that has four major fitness peaks, one of which includes the SGC. Furthermore, genetic algorithm analysis revealed that evolution under such a multi-peaked fitness landscape could be strongly biased toward a narrow peak, in an evolutionary path-dependent manner. 
\end{abstract}
\maketitle

\section{Introduction}
The genetic code is the rule encoding the mapping between 64 codons and 20 different amino acids or a terminal codon. The standard genetic code (SGC), shared by almost all organisms, has an apparently non-random structure~\cite{woese1965evolution}, in which the same amino acid is coded blockwise across adjacent codons in the codon table (Fig.~1a; a block corresponds to a set of codons coding the identical amino acid), and is thus thought to be subject to selection pressure. 
Based on this, Woese~\cite{woese1965evolution} proposed the error-minimization hypothesis, which postulates that the SGC is selected to enhance robustness against translational (or possibly mutational) errors. Several subsequent theoretical studies, comparing the SGC to hypothetical random genetic codes in terms of robustness, support this hypothesis~\cite{haig1991quantitative, freeland1998genetic, gilis2001optimality, goodarzi2004optimality,archetti2004codon}.
These studies estimated robustness using a simple formula ~\cite{haig1991quantitative, freeland1998genetic, gilis2001optimality, goodarzi2004optimality,archetti2004codon, shenhav2020resource} that can apply to any given genetic code, and the estimated robustness can be regarded as the genetic code fitness. 
By estimating the fitness for multiple possible codes, this formula in principle enables analysis of the fitness landscape of the genetic code. The fitness landscape is a useful concept to describe evolutionary dynamics since evolution can be viewed as an adaptive walk on the landscape toward a higher-fitness region. The analysis of the fitness landscape of the genetic code would thus provide a lot of information about how the genetic code has evolved to the SGC and whether the evolution to the current SGC was inevitable, or whether it could have evolved to a completely different genetic code.
Indeed, by estimating how strongly the SGC is selected with respect to translational robustness, some theoretical studies~\cite{haig1991quantitative, freeland1998genetic, gilis2001optimality, goodarzi2004optimality,archetti2004codon,novozhilov2007evolution,koonin2009origin} have shown that a huge number of local fitness peaks exist~\cite{novozhilov2007evolution}, suggesting that the landscape is extremely rugged, and the SGC is located near one of these local peaks. 
However, these studies have been restricted to exploring a rather limited repertory of genetic codes due to high computational cost and thus analyzing only the local structure of the landscape around the SGC.
Because of this, the code evolution to the current SGC has not yet been clarified, as well as whether the high fitness codes all have a similar structure to the SGC or can have a completely different structure from the SGC. 
 To elucidate these, the global structure of the fitness landscape needs to be analyzed. 
 A single prominent peak with many small local peaks in the global structure of the fitness landscape would lead one to conclude that all high-fitness codes have a structure similar to that of the SGC. On the other hand, if there are multiple prominent peaks, each with many local peaks, it means that the current SGC occurrence is not a unique solution, but contains some coincidence.

A genetic code can be represented by a table with 64 entries, in which each entry corresponds to a single codon, a sequence of three DNA bases (A, G, C, T) or RNA bases (A, G, C, U). The genetic code maps each codon to one of 20 different amino acids or to a terminal signal (Fig.~1a). Here, the code $\ve{a}$ is a 64-dimensional vector, where each entry $a(c)$ for $c = 1(UUU), ..., 64(GGG)$ takes one of 20 different amino acids or a terminal signal. If a translation error causes codon $c$ to be mistakenly recognized as $c'$, an incorrect amino acid $a(c')$ is introduced into the protein. This error would cause an associated fitness reduction due to protein misfolding or malfunction. It is natural to assume that the reduction in fitness depends on the difference in physicochemical properties $d(a(c),a(c'))$ between amino acids $a(c)$ and $a(c')$. 
Given that hydrophobic interaction is the dominant driving force in protein folding, this difference in physicochemical properties can reasonably be represented as a difference in hydrophobicity or hydrophilicity.
Indeed, Haig and Hurst~\cite{haig1991quantitative} examined four amino acid's properties (polar requirement, hydropathy, molecular volume, and isoelectric point) and suggested that the SGC has been under selection pressure to minimize changes due to misreading in the polar requirement scale~\cite{woese1966fundamental}, a measure associated with hydrophilicity.
Following these studies, we adopted $d(a(c),a(c'))$ as the square of the difference in the polar-requirement scale between $a(c)$ and $a(c')$. The polar-requirement values of each amino acid are illustrated by the colors in Fig.~1(a), and described in the Supplemental Materials.
The average cost of misreading a codon associated with code $\ve{a}$ is given by 
\begin{equation} \label{eq:cost}
cost(\ve{a})=\sum_c \sum_{c'}P(c'|c)d(a(c),a(c')),
\end{equation}
where $P(c'|c)$ is the probability of misreading $c$ as $c'$. 
\begin{figure*}
\centering
\includegraphics[width=16cm]{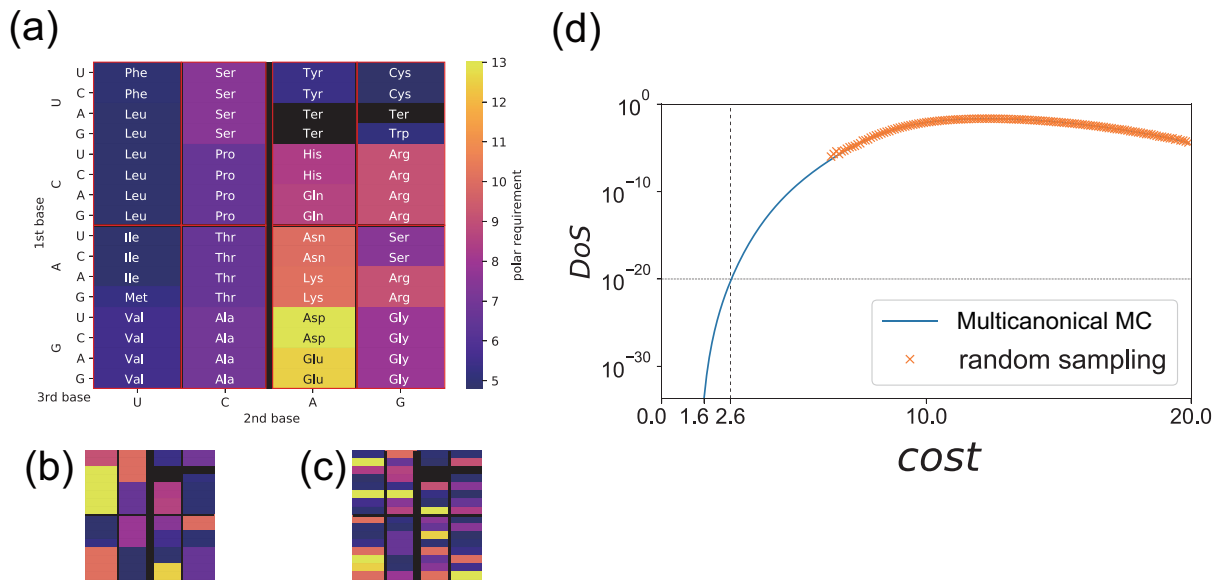}
\\
\begin{flushleft}
\caption{{\bf The standard genetic code and randomly generated genetic codes:}} (a) The standard genetic code. 
(b) A random genetic code generated from a random ensemble that Haig and Hurst~\cite{haig1991quantitative} adopted. The block structure in the SGC where the same amino acid is coded is constrained to be maintained in the randomly generated code (see also Fig.~S2). (c) A random code generated from a fully random ensemble where each codon maps to any one of 20 amino acids. (d) Cost density-of-states function for the proposed random genetic code ensemble. The blue line represents the estimates by the multicanonical Monte Carlo (multicanonical MC). The orange crosses indicate estimates based on naive random sampling.
 \label{fig:fig1}
\end{flushleft}
\end{figure*}

The calculation of the cost defined by Eq. ~\ref{eq:cost} for randomly generated codes provides insights into how evolutionarily optimized the SGC is, in terms of robustness against translational error, and how unlikely it is that the current SGC evolved by coincidence (i.e., without selection).
In their seminal paper, Freeland and Hurst~\cite{freeland1998genetic} generated random genetic codes using a random sampling method, estimated the fraction of codes with lower costs (i.e., higher robustness against translational error) than the SGC, and concluded that the probability of obtaining equally or more robust genetic code is $\sim 10^{-6}$. 
Although these studies provide an excellent basis for understanding genetic code evolution, they were limited in their search space due to the high computational cost and considered only a random ensemble of codes maintaining the codon block-structure of the SGC, in which a block corresponds to a set of codons coding the identical amino acid (Fig.~1b and Fig.~S2). 
Most of the studies in this field have considered similar random genetic code ensembles using limited search spaces. Such studies have examined, for instance, evolutionary simulation of the genetic code~\cite{novozhilov2007evolution}, and methods to improve the cost function to incorporate protein amino-acid frequency~\cite{gilis2001optimality,zhu2003codon}.
However, this choice of random genetic code ensemble is inconsistent with examples of the non-standard genetic code~\cite{sengupta2015pathways,jukes1993evolutionary,yokobori2010evolution}, whereby the code differs in only a single or a few codons from the SGC, altering the block structure of the SGC. 
A justifiable choice of the random genetic code ensemble is such that each codon can take 20 possible amino acids, with the constraint that the whole code contains 20 different amino acids (Fig.~1c); thus, the number of possible codes in the ensemble is approximately $20^{64} \sim 10^{83}$, which is much larger than the number of possible codes in the ensemble ($20! \sim 10^{18}$) maintaining the same block structure of the SGC.
However, a fully random ensemble with $20^{64}$ states produces a 64-dimension search space. Given this size of the search space, naive random sampling, in which all random genetic codes are generated with equal probability, cannot be used to obtain a meaningful genetic code with a cost as low as that of the SGC.

Here, we applied multicanonical Monte Carlo (MC) analysis, which can effectively sample low-cost genetic codes even from an ensemble of fully random genetic codes, to reveal the global structure of the fitness landscape. The resulting fitness landscape had four major peaks, one of which includes the SGC. This indicates that there are four classes of high-fitness genetic code and suggests that the current genetic code potentially could have been the other three classes. 

\section{Method}
For the computation of Eq.(1), we adopted the following $P(c'|c)$ according to previous studies~\cite{freeland1998genetic, gilis2001optimality, goodarzi2004optimality,archetti2004codon,zhu2003codon} that are based on experimental observation~\cite{friedman1964lack}:
\begin{eqnarray}
P(c'|c)  \propto \left\{
\begin{array}{cc}
1 & \mbox{for a change in the 3rd base }\\
1 & \mbox{for a change in the 1st base by TS}\\
0.5 & \mbox{for a change in the 1st base by TV}\\
0.5 & \mbox{for a change in the 2nd base by TS}\\
0.1 & \mbox{for a change in the 2nd base by TV} \\
0 & \mbox{double/triple base changes}
\end{array}
\right. ,
\end{eqnarray}
where TS represents a transition [an interchange between purines (A $\leftrightarrow$ G) or between pyrimidines (C $\leftrightarrow$ U)], and TV denotes a transversion [an interchange of purines for pyrimidines (A $\leftrightarrow$ C, A $\leftrightarrow$ U, G $\leftrightarrow$ C, G $\leftrightarrow$ U)]. 
This $P(c'|c)$ indicates that the most frequent errors are mistranslations within the red box in Fig.~1a, while misreading of a codon into a codon in an adjacent red box, without crossing the black thick line at the center, is half as likely, and the misreading of a codon by crossing the center line is 10 times less likely (Fig.~S1).
For misrecognition of an amino acid as the terminal signal (where $a(c')$ is the terminal signal), a cost $c_0$, independent of the polar requirement of $a(c)$, is assigned. Likewise, for misreading the terminal signal as an amino acid (that is, $a(c)$ is the terminal signal) a constant cost $c_1$ is assigned.
As the choice of $c_0$ and $c_1$ affects only the basal value of the cost, and does not alter the cost difference between any two codes, we adopted $c_0=c_1=0$. Only a single base change is considered here, because the probability of double or triple base changes are negligibly small.

For sampling from such an ensemble, we apply a computational technique termed as rare event sampling using multicanonical Monte Carlo~\cite{berg1991multicanonical,berg1992multicanonical}. Multicanonical Monte Carlo has been developed in statistical physics such as spin systems ~\cite{wang2001efficient, berg1991multicanonical,berg1992multicanonical} and protein folding~\cite{chikenji1999multi, higo2012enhanced}. This method enables efficient sampling of rare events, exemplified by low-energy states, without becoming trapped in local minima. Furthermore, this method provides an estimate of the density-of-state, which represents the energy distribution for all possible states in the system. This method combines the advantages of an optimization algorithm (energy minimization) and naive random sampling (calculation of energy distribution), and has been applied to a variety of fields other than physics, such as random matrix theory~\cite{saito2010multicanonical}, combinatorial optimization~\cite{kitajima2015numerous}, and the generation of surrogate time-series data~\cite{iba2014multicanonical}. 
Recently, this method has also been applied to gene regulatory network evolution~\cite{nagata2020emergence, kaneko2022evolution, saito2013robustness}.

Here, we consider the Markov chain sampling of random genetic code to be $\ve{a} \to \ve{a'} \to \cdots$, where a candidate state $\ve{a'}$ is generated from $\ve{a}$ by flipping the $c$-th index of $\ve{a}$, $a(c)$, to another amino acid $a'(c)$ [$\neq a(c)$]. The transition $\ve{a} \to \ve{a'}$ is determined by accepting or rejecting the candidate state $\ve{a'}$ of the transition probability function, $\Pi_{cost(\ve{a}) \to cost(\ve{a'})}=\mbox{min}(1,w(cost(\ve{a'}))/w(cost(\ve{a}))$, where $w$ is the ``weight'' function (a function of the cost). The sampled distribution of the cost, $P_{eq}(cost)$, obeys the following detailed balance condition for a large sample size:
\begin{eqnarray}
\Pi_{cost \to cost'} P_{eq} (cost) = \Pi_{cost' \to cost} P_{eq} (cost'),
\end{eqnarray}
and thus depends crucially on the weight. While the Gibbs distribution of $P_{eq}(cost)$ is obtained for the choice of an exponential function of $cost$ in ordinary Metropolis sampling, multicanonical sampling adopts the multicanonical weight $w(cost) \propto 1/\Omega(cost)$, where $\Omega(cost)$ is the density-of-state, leading to a uniform distribution of $P_{eq} (cost)$, since $P_{eq} (cost) \propto \sum_{\ve{a}} w(cost(\ve{a})) I(cost(\ve{a})=cost) = w(cost)\Omega(cost)$, where $I(X=Y)$ is the function that takes I(X=Y) = 1 if the equality is satisfied, and zero otherwise. Therefore, for multicanonical sampling, the sequence of states $\ve{a}$ generated by the Markov chain is regarded as a random walk in cost space, ensuring efficient sampling of rare events (i.e., low cost states) without becoming trapped in local minima. Since $\Omega (cost)$ is not known a priori, one needs to numerically compose $w(cost)$ that gives rise to a flat sampling histogram of the cost space. Here, we used the Wang-Landau algorithm~\cite{wang2001efficient} to construct and tune the weight function $w(cost)$, subsequently performing multicanonical sampling with a fixed $w(cost)$~\cite{iba2014multicanonical}.

\section{Results}
First, we computed the fraction of genetic codes having lower cost than the SGC in a random genetic code ensemble. To avoid generating biologically meaningless genetic codes, such as those comprising only the terminal signal, which would obviously have the lowest cost, the numbers and positions of terminal signals were set to be the same as for the SGC. Similarly, 
we imposed the restriction that any code $\ve{a}$ should contain 20 different amino acids. 
In addition, we required that any code in the ensemble contain at least as many aspartic acids (Asp) and glutamates (Glu) as the SGC (i.e., at least two of each), because they have the highest polar-requirement values, and the cost value depends critically on their numbers. Without this requirement, almost all ($\sim$ 99\%) of the sampled genetic codes with low cost contain only a single Asp and Glu, which is not compatible with our goal of visualizing the fitness landscape to which the SGC belongs (Supplemental Materials and Fig.~S6). 
The total number of possible genetic codes in this ensemble was $\sim 10^{79}$, which is still much larger than that reported in previous studies~\cite{freeland1998genetic, gilis2001optimality, goodarzi2004optimality,archetti2004codon,zhu2003codon,novozhilov2007evolution}.
Naive random sampling with a sample size $10^{10}$ could not generate a code with a lower cost than the SGC (Fig.~1d). We performed multicanonical sampling with $10^{11}$ steps by fixing the weight function $w(cost)$.
The density of states $\Omega (cost)$ is obtained by $\Omega (cost) \propto H(cost)/w(cost)$, where $H(cost)$ is the sampling histogram of the cost. 
The estimates of $\Omega (cost)$ (Fig.~1d) were consistent with the naive random sampling results for a large cost, revealing that 
the density of genetic codes having a cost lower than that of the SGC is ca. $10^{-20}$. Such a genetic code is virtually impossible to generate via naive random sampling.
This estimate is much smaller than the previous estimates, $\sim 10^{-6}$, indicating that the SGC is much more evolutionarily optimized for robustness against errors than previously thought.

\begin{figure*}
\centering
\includegraphics[width=16cm,pagebox=cropbox,clip]{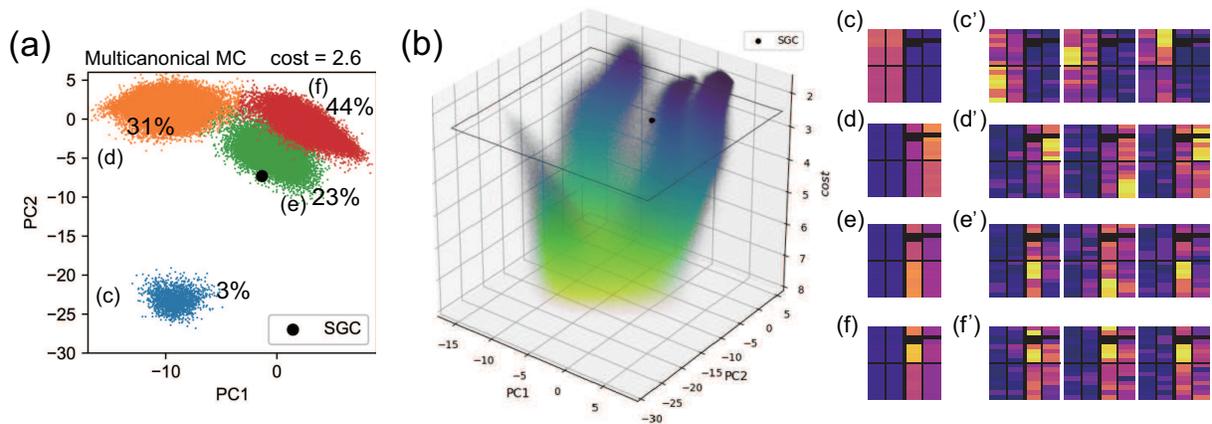}
\\
\begin{flushleft}
\caption{{\bf Fitness landscape under a random genetic code ensemble.}} (a) Genetic codes with comparable cost values to the SGC, where $2.6-\Delta<cost <2.6+\Delta$ ($\Delta=1.0$) in the PC1-PC2 space (the contribution ratios: PC1, 27\%, PC2, 14\%). Principal component analysis and k-means clustering were performed using the sampled data with a cost value of $cost <2.6+\Delta$. (b) Fitness landscape visualized by piling up the scatterplot on the PC1-PC2 plane, with costs decreasing on the vertical axis, where low cost corresponds to high fitness. (c-f) Average structure of the genetic code in each cluster. (c'-f') examples of samples belonging to the clusters (c-f) in (a). 
\end{flushleft}
\end{figure*}

Next, we investigated the characteristics of genetic codes with low cost (i.e., high fitness). Using the sampled data with costs comparable to, or lower than, those of the SGC (i.e., $cost < 2.6 + \Delta$, where $\Delta=1$), we reduced the 64-dimensional code space to two-dimensional space, via principal component analysis (PCA). PCA was performed for the set of 64-dimensional vectors $\ve{a}$, in which each entry contains the polar-requirement scale of the encoded amino acid at the codon.
The resultant scatter plot in the PC1-PC2 space for $cost=2.6$ $\pm \Delta$ (the same cost as the SGC) (Fig.~2a) reveals that there are four clusters, among which the data are unevenly distributed ($\sim$ 44\% in the red cluster, $\sim$ 31\% in the orange, $\sim$ 23\% in the green, and $\sim$ 3\% in the blue). To determine the number of clusters, we used the k-means method with the elbow method (Fig.~S3).
By piling up a two-dimensional scatterplot of different cost values on the vertical axis, we can visualize the cost landscape, the counterpart of the fitness landscape, where low cost corresponds to high fitness, and vice versa. 
Figure~2(b) clearly shows the multimodality of the fitness landscape, where the four-peak structures in the low-cost region are connected to the high-cost region.
Figs.~2(c-f) illustrate the average composition of the genetic codes within each peak for a genetic code with the same cost as the SGC ($cost=2.6$), and Figs.~2(c'-f') show three representative examples in each cluster. 
Comparing Fig.~2(a) and (c-f), we see that a genetic code with high PC1 values tends to have high polar-requirement amino acids in the third column, whereas the larger the value on PC2, the more high polar-requirement amino acids tend to be coded in the third and fourth columns (the right half of the codon table). 
The green cluster in Fig.~2(a), to which the SGC belongs, was further analyzed via PCA of the dataset belonging to the cluster. This revealed a similar genetic code structure to the SGC (Fig.~S4).

\begin{figure*}[!t]
\centering
\includegraphics[width=14cm,pagebox=cropbox,clip]{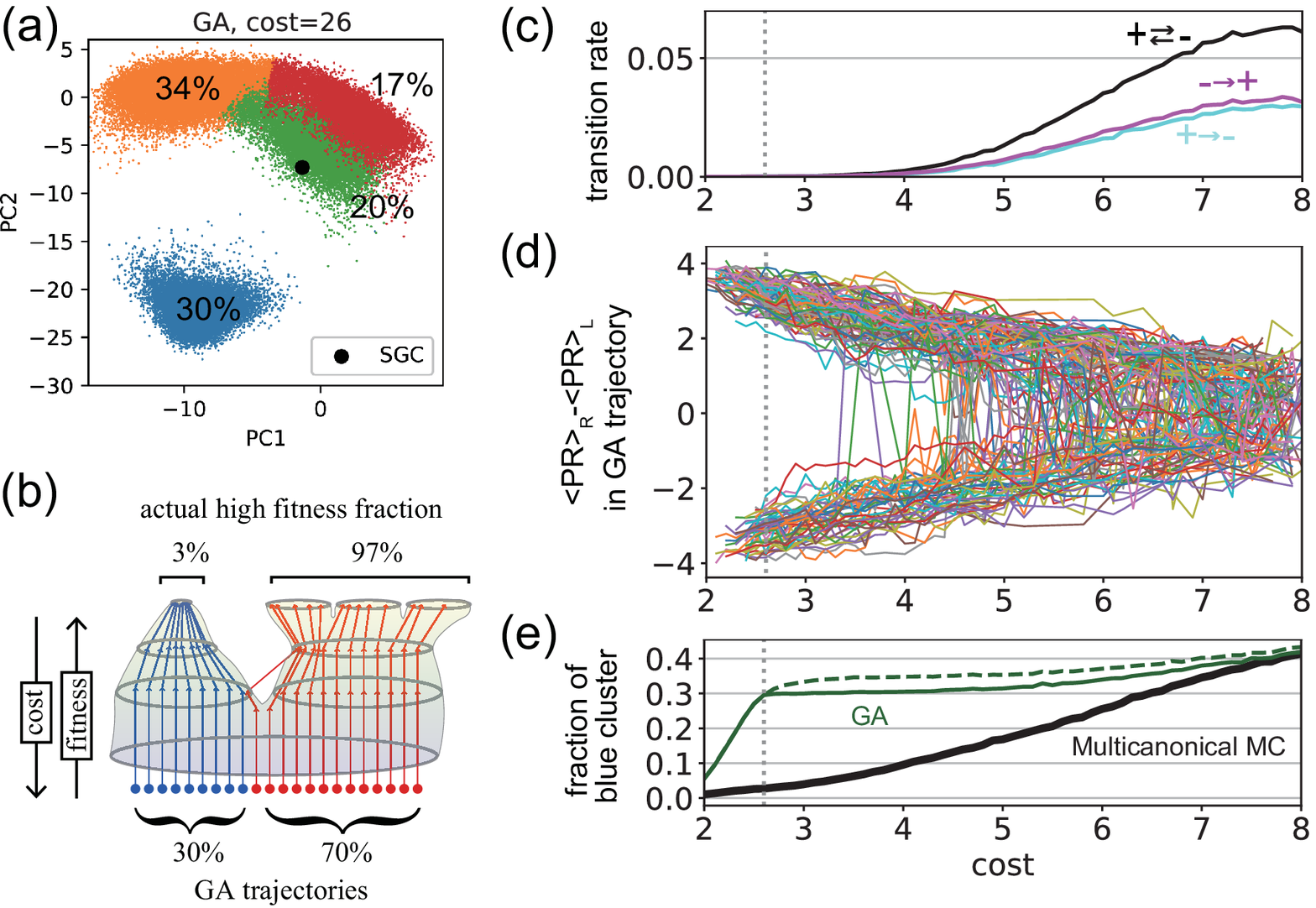}
\\
\begin{flushleft}
\caption{{\bf Simulation results by genetic algorithms (GAs):}} \label{fig:GA}
Results of $10^5$ independent GAs with population size $N=100$ and $500$ generations, where the top $50\%$ of the population is selected in each generation. (a) Codes with comparable cost values to the SGC, $2.6-\Delta<cost <2.6+\Delta$ ($\Delta=1.0$), on the same PC1-PC2 plane as in Fig.~2. Clustering was performed based on samples with $cost <2.6+\Delta$, obtained via GA and MCMC sampling.
(b) Schematic illustration of evolution under the multipeak fitness landscape. 
(c-e) Analysis of the independent GA trajectories. The cost values of the independent GA runs are plotted on the horizontal axis. (c) Transition rates from $\Delta_{PR}>0$ to $\Delta_{PR}<0$ (black), from $\Delta_{PR}<0$ to $\Delta_{PR}>0$ (purple), and for the inverse transition (cyan). 
(d)$\Delta_{PR}$ in GA trajectories as a function of the cost value experienced in each GA. Note that if a trajectory experiences the same cost value multiple times, only the latest value is recorded, so a single curve is always a univalent function and does not necessarily correspond to the GA time series (GA time series can go up and down in cost). 
(e) Fraction of genetic codes with $\Delta_{PR}<0$. The green dashed line depicts the fraction for all the GA trajectories, while the thick line indicates that for GA trajectories conditioned to reach cost=2.6. 
\end{flushleft}
\end{figure*}
Our use of multicanonical MC enabled unbiased sampling of random genetic codes, enabling us to analyze and visualize the global structure of the genetic code fitness landscape.
In the reconstructed fitness landscape, the optimized codes were categorized into four clusters (Fig.~2a, b): the blue cluster, in which highly polar amino acids are located in the two left columns, occupies only a small fraction ($\sim 3\%$, for $cost < 2.6+\Delta$).
Interestingly, genetic algorithms (GAs), widely used evolutionary simulations, can cause large deviations from unbiased sampling, because the evolutionary outcome could be strongly biased in an evolutionary path-dependent manner.
To demonstrate this, we performed a GA in which the population of genetic codes was selected according to the cost in Eq.~(\ref{eq:cost}) and copies of the selected population are generated by random sampling without replacement. In these copies, mutations in the genetic code (i.e., from $a(c)$ to another amino acid, $a'(c)$) occur with a mutation rate $\mu$ in each element of $\ve{a}$. When the mutated code does not contain 20 different amino acids, or at least two instances of Asp and Glu, the mutation process starts over. We performed $10^5$ independent GA runs. 
 
Fig.~\ref{fig:GA}a illustrates the resultant genetic codes that reached a cost of 2.6 (the same cost as the SGC) on the same PCA projection as in Fig.~2a; this further illustrates that the GA can reach all four peaks, with up to 30\% of the GA results concentrated in the blue cluster. However, the actual phase volume of the cluster occupies only 3\% of the high-fitness code space (Fig.~2a, blue cluster). To understand how this bias arises, we analyzed the optimization trajectories of the independent GA runs (Fig.~\ref{fig:GA}b-e). The blue cluster in Fig.~\ref{fig:GA}a comprises codes where high polar-requirement amino acids are located in the two left columns (see also Fig.~2c and c'); the tendency to approach the cluster via a GA trajectory is characterized by $\Delta_{PR}$, the difference between the average polar-requirement value in the two right columns $\langle PR \rangle_R$ and the two left columns $\langle PR \rangle_L$ of the code. 
Figure~\ref{fig:GA} (d) illustrates that $\Delta_{PR}=\langle PR \rangle_R - \langle PR \rangle_L$ through GA trajectories as a function of the cost values experienced in a single GA run, where genetic codes with negative $\Delta_{PR}$ values around $cost=2.6$ represent the blue cluster in Fig.~\ref{fig:GA}a.
In Fig.~\ref{fig:GA}c, the number of transitions between positive and negative $\Delta_{PR}$ declines as the optimization proceeds, with almost none occurring below $cost = 4 \sim 5$.
Consistent with this, the fraction of genetic codes with $\Delta_{PR}<0$ appearing in the GA optimization trajectories exhibits a plateau below $cost = 4 \sim 5$ (Fig.~\ref{fig:GA}e: thick green line), whereas the actual fraction of codes estimated via multicanonical MC (Fig.~\ref{fig:GA}e: thick black line) declines monotonically as cost decreases. This indicates that the fraction of codes with a negative $\Delta_{PR}$ below $cost\simeq 2.6$ (i.e., the blue cluster in Fig.~\ref{fig:GA}a) is determined early in the optimization process, at $cost>5$, being
determined by the phase-space volume with $cost>5$ rather than $cost\simeq 2.6$. 
This leads us to conclude that the GA can cause a large bias in the situation depicted in Fig.~\ref{fig:GA}b, where the fitness landscape has multiple peaks that allow no peak-to-peak jump, and one peak (the left peak in Fig.~\ref{fig:GA}b) has a broad waist and narrow tip. 
In such a situation, the fraction of trajectories reaching the left peak is determined by the waist size of the fitness landscape; in other words, the basin size of the peak, when the optimization dynamics are interpreted as a dynamic system, causes the overconcentration in the left narrow peak.

\section{Discussion}

The proposed multicanonical Monte Carlo method, which efficiently samples rare events in a fully random genetic code ensemble, revealed that the probability of obtaining a genetic code with a cost as low as that of the SGC by chance is ca. $10^{-20}$, much lower than the previous estimate~\cite{freeland1998genetic}. It would be virtually impossible to generate such a code by naive random sampling.
The proposed multicanonical method would therefore be appropriate as a standard tool in studies inferring the selection pressure on the SGC, instead of the previously used naive random sampling method
~\cite{freeland1998genetic, gilis2001optimality, goodarzi2004optimality,archetti2004codon,zhu2003codon,novozhilov2007evolution,shenhav2020resource}.
Note that despite their extremely low probability, the number of such codes was huge as $\sim 10^{59}$. 
Despite their extremely low probability, the number of such codes was $\sim 10^{59}$. This ``numerous but rare'' problem is also found in the ``magic square'' problem~\cite{kitajima2015numerous}.

The genetic code fitness landscape had four peaks, one of which included the SGC, indicating that the genetic code might have evolved as one of the other three classes. Further, we found that, when applying the GA, evolutionary path-dependent bias arises under the multi-peaked fitness landscape, where peak-to-peak jumps are not allowed within a finite number of generations. Under such conditions, the fraction of independently evolved genetic codes reaching each peak is determined by the basin volume rather than the phase-space volume of these optima, when the evolutionary dynamics is interpreted as a dynamical system maximizing fitness. From the perspective of statistical physics, this fraction is dictated by the sum of the path probabilities from a random initial condition to each peak, rather than the Boltzmann distribution of the fitness function; the fraction should coincide with the Boltzmann distribution when the system is in equilibrium but can differ in a finite time.
In the proposed scenario, the evolutionary outcome is determined by the ease of finding a peak of the fitness landscape, but not by the phase-space volume of optimal genotypes. This scenario is applicable to both evolutionary studies and optimization problems with gradient-descent-like algorithms under a multi-peaked fitness landscape. A bias specific to the evolutionary algorithm has also been reported in a study of gene regulatory networks ~\cite{kaneko2022evolution}, in which the mutational robustness was significantly higher in the GA than in the reference ensemble at the same fitness level. It remains to be clarified whether this bias toward mutational robustness~\cite{kaneko2022evolution,wagner2005robustness} is related to the bias caused by the presence of multiple peaks, presenting a new avenue for evolutionary studies.

The asymmetric four-peak shape of the fitness landscape (Fig.~2b) is attributed to our assumption that the evolution of the terminal signal is ignored. However, this does not mean that the results obtained are mere artifacts with no meaning to interpret, nor that our assumptions are unrealistic.
An asymmetric fitness landscape (e.g., Fig.~2b) emerges naturally when the terminal codons are localized on the table to reduce the cost, as in the case of SGC. 
In this sense, the fitness landscape obtained in this study should be considered as a landscape under the conditional probability that the terminal signals are in the same position as in the SGC.
The following aspects that were beyond the scope of our study require future analysis: the incorporation of the effect of codon usage~\cite{zhu2003codon,archetti2004codon}, coevolution of the genetic code with the amino-acid synthesis metabolic pathway~\cite{higgs2009four,wong1975co,koonin2009origin}, and the recently reported robustness of the genetic code against mutations that increase carbon and nitrogen incorporation~\cite{shenhav2020resource}.
In particular, our analysis did not answer why nearly all organisms use the same genetic code, the SGC, instead of using different codes for different organisms.
The emergence of a universal code could be explained by the effects of horizontal gene transfer (HGT), as Vetsigian et al.~\cite{vetsigian2006collective} addressed. How the HGT can alter the evolutionary outcome under the proposed fitness landscape will also be one of the issues to be addressed. 
Our proposed method has a flexibility to incorporating several factors as listed above and would provide a promising means of estimating the rareness of SGC from a fully random genetic code ensemble, and visualizing the fitness landscape of the code.

\section*{Acknowledgments}
We would like to acknowledge the helpful discussions with Yukito Iba, Macoto Kikuchi and Kunihiko Kaneko.
This study was supported in part by the Japan Society for Promotion of Science (JSPS) KAKENHI(17H06389 to CF), Japan Science and Technology Agency (JST) ERATO (JPMJER1902 to CF), and Cooperative Study Program of Exploratory Research Center on Life and Living Systems (ExCELLS; program No. 20-102, 21-102 to N.S.)

\bibliographystyle{unsrt}

\end{document}